\documentclass{article}

\usepackage{multirow}
\usepackage{PRIMEarxiv}

\usepackage[utf8]{inputenc} 
\usepackage[T1]{fontenc}    
\usepackage{hyperref}       
\usepackage{url}            
\usepackage{booktabs}       
\usepackage{amsfonts}       
\usepackage{nicefrac}       
\usepackage{microtype}      
\usepackage{lipsum}
\usepackage{fancyhdr}       
\usepackage{graphicx}       
\graphicspath{{media/}}     
\usepackage{amsmath}
\usepackage{newfloat}
\DeclareFloatingEnvironment[name={Supplementary Figure}]{suppfigure}

\title{A Metaheuristic for Amortized Search in High-Dimensional Parameter Spaces
}

\author{
  Dominic Boutet \& Sylvain Baillet \\
  Montreal Neurological Institute \\
  McGill University \\
  Montreal QC, Canada \\
  \texttt{\{Dominic Boutet\}}dominic.boutet@mail.mcgill.ca \\
  \texttt{\{Sylvain Baillet\}}sylvain.baillet@mcgill.ca 
}

\begin{document}
\maketitle

\begin{abstract}
Parameter inference for dynamical models of (bio)physical systems remains a challenging problem. Intractable gradients, high-dimensional spaces, and non-linear model functions are typically problematic without large computational budgets. A recent body of work in that area has focused on Bayesian inference methods, which consider parameters under their statistical distributions and therefore, do not derive point estimates of optimal parameter values. Here we propose a new metaheuristic that drives dimensionality reductions from feature-informed transformations (DR-FFIT) to address these bottlenecks. DR-FFIT implements an efficient sampling strategy that facilitates a gradient-free parameter search in high-dimensional spaces. We use artificial neural networks to obtain differentiable proxies for the model's features of interest. The resulting gradients enable the estimation of a local active subspace of the model within a defined sampling region. This approach enables efficient dimensionality reductions of highly non-linear search spaces at a low computational cost. Our test data show that DR-FFIT boosts the performances of random-search and simulated-annealing against well-established metaheuristics, and improves the goodness-of-fit of the model, all within contained run-time costs.
\end{abstract}


\newpage
\section{Introduction}
Dynamical models are ubiquitous in applied sciences and engineering. Leveraging such models in conjunction with empirical data requires performing inference on the model parameter space \cite{cranmer_frontier_2020}. This is typically approached by specifying an objective function, defined in terms of the inference objective, and solved using optimization algorithms \cite{villaverde_benchmarking_2019}.

Large-scale and/or non-linear models challenge this model-inference approach because of practical considerations, whereby parameter optimization cannot be derived analytically, and/or direct applications of classical gradient-based techniques may be intractable \cite{schalte_evaluation_2018}. For this reason, gradient-free methods have been explored as alternatives  \cite{schalte_evaluation_2018}, typically operating iteratively through successive search and update steps \cite{schalte_evaluation_2018}. The outcome of the resulting inference procedure can be an empirically-derived parameter distribution or the associated likelihood, over the parameter space or a set of point estimates with specific parameter values \cite{cranmer_frontier_2020}.

However, such inference approaches are susceptible to under-perform when model behavior is overly sensitive within certain regions of the parameter space \cite{deistler_energy-efficient_2022}. Further, because these methods span the full parameter space, they do not provide insight into the model dynamics that would be useful to, e.g., in-silico experiments based on biophysical models, or other intensive studies based on model simulations, which require the derivation of point estimates via, e.g., maximum a posteriori estimation (MAP) \cite{cranmer_frontier_2020}. Both the inference and MAP steps induce cumulative errors and add to computational costs, which negatively impacts practicality. Search algorithms such as random-search (RS), simulated-annealing (SA), particle-swarm-optimization (PSO) and covariance-matrix-evolutionary-strategies (CMA-ES) are alternatives to inference-based parameter optimization. They proceed by exploring the parameter space to find point estimates that optimize the objective function \cite{karnopp_random_1963,yang_chapter_2021-1,yang_chapter_2021,jastrebski_improving_2006}. However, these approaches do not scale efficiently to high-dimensional parameter spaces \cite{schalte_evaluation_2018}. Further, we are not aware of a search metaheuristic for the amortization of the computational load of the point estimation type of inference problem in cases of multiple empirical observations. 

To address these limitations, we introduce an amortized metaheuristic for search algorithms that is computationally efficient and scales to high-dimensional non-linear problems. The dimensionality reductions from feature-informed transformations (DR-FFIT) approach improves the efficiency of parameter search using neural subspace estimators (NSE), which are neural networks trained to estimate a locally sensitive lower dimensional search space within the given sampling region. NSEs first learn a mapping between model parameters and predefined features of the model outputs. This mapping is then used to derive local parameter sensitivity within the specified sampling region of the parameter space. The features may not be specific to empirical observations, therefore a trained NSE may be used in the context of multiple empirical observations, resulting in an amortization of the computational cost.

Below we test the DR-FFIT metaheuristic on two use-cases inspired by computational neuroscience modeling problems with inference performed on multiple observations per model. The first test case  involves the Wilson-Cowan neural mass model used in conjunction with empirical observations of power spectrum densities (PSD) derived from human resting-state magnetoencephalography \cite{wilson_mathematical_1973,baillet_magnetoencephalography_2017, niso_omega_2016}. The second case involves a simplified Hodgkin-Huxley current clamp neuron model in conjunction with simulated time series of neuronal membrane potentials \cite{hodgkin_quantitative_1952,pospischil_minimal_2008}.

\section{Material \& Methods}
\label{sec:headings}
We first establish the notation and terminology. We then describe a general framework of parameter search. Finally, we introduce the DR-FFIT metaheuristic. 

\subsection{Notation \& terminology}
Let us define a dynamical system:
\begin{equation}
\dot{x}=f(x,t;\theta, x_0) + \sigma,
\end{equation}
where function $f$ defines the dynamics of the system, $x$ is the vector of state variables, $t$ the time variable, $\theta$ the set of model parameters from the parameter space $\Theta$, $x_0$ the state vector of initial conditions, and $\sigma$ representing additive noise. Let us now define $Y_C = \{Y_0, ..., Y_c\}$ as a set of distinct empirical observations, e.g., the power spectrum densities derived from the MEG data. We then define the model prediction
\begin{equation}
Y_{\theta}' = \Psi(\theta),
\end{equation}
as an numerical approximation of the model output by a numerical simulator $\Psi$ for the parameter set $\theta$. Finally, we define the objective functions $L_i(\theta)$ for each observation $Y_i \in Y_C$ via a loss function $\ell$ (e.g., mean-squared error):
\begin{equation}
L_i(\theta) = \ell(Y_{i},Y_{\theta}').
\end{equation}

\subsection{Parameter search and metaheuristics}
Parameter search methods explore the parameter space $\Theta$ following a sampling distribution $D^j_{\Theta}$ updated after each iteration. These algorithms can be summarized, for a given objective function $L_i$ at iteration $j$, as:
\begin{enumerate}
    \item Draw a set of $N$ samples from the sampling distribution: $\theta_N^j = \theta_0^j, \theta_1^j, ..., \theta_n^j \sim D_{\Theta}^j$, 
    \item Compute $L_i(\theta_N^j)$ and designate $\theta_{c}^j = \text{argmin}_{\theta} L_i(\theta_N^j)$ as the best solution from the set of samples,  
    \item Update the estimate $\hat{\theta_i} \leftarrow \theta_c^j$ if $L_i(\theta_c^j) < L_i(\hat{\theta_i})$, and update the sampling distribution $D_{\Theta}^{j+1}$.
\end{enumerate}
Metaheuristics differ in terms of type of sampling distribution used and how it is updated at each iteration. In the present study, RS uses a uniform hyper-sphere centered around $\hat{\theta_i}$ and $D_{\Theta}$ is updated with new $\hat{\theta_i}$ values \cite{karnopp_random_1963}. Here, we used the performances obtained with RS as a baseline to the other tested approaches. We implemented SA as a variation of RS with an exponentially decaying exploration-to-exploitation ratio indexed by the annealing temperature. SA occasionally selects $\theta_c^j$ instead of $\hat{\theta_i}$ to update $D_{\Theta}$ based on the acceptance probability \cite{yang_chapter_2021-1}. Within this framework, PSO is conceived as sampling from a distribution defined over the possible future locations of particles in a subset of the parameter space. These locations are updated from their current values by applying a random velocity vector \cite{yang_chapter_2021}. Finally, CMA-ES relies on a multivariate normal sampling distribution that is updated using a parameter covariance matrix learned through an evolutionary strategy\cite{jastrebski_improving_2006}. Here we used the CMA-ES implementation available from the pycma package \cite{hansen_cma-espycma_2023}.

\subsection{The DR-FFIT metaheuristic}
Let us define an arbitrary feature space $W$ for the model predictions, such that a feature function $F$ is defined to map a model prediction $Y_{\theta}'$ to feature $w_{\theta}$ for a given parameter $\theta$: $w_{\theta} = F(Y_{\theta}')$. $F$ can be pre-defined or learned from data (e.g., with a principal component analysis, PCA). 
Let $T: \Theta \rightarrow W$ be a function that directly maps the parameter space onto the feature space such that $T(\theta) = w_{\theta}$ (Figure 1). In essence, $T$ models the direct relationship between the parameters and the features, therefore the gradient of $T$ carries information about the sensitivity of model parameters to model features. The DR-FFIT metaheuristic is based on a neural subspace estimator (NSE), which is a neural network(s) $T'$ trained to approximate $T$ from model simulations. Neural networks are differentiable, therefore this approach enables a computationally efficient approximation of the gradient of $T$ across the parameter space. The NSE can then leverage $\nabla T'$, computed using backpropagation over a set $\theta_{M}$ of $m$ points within the sampling region ($\theta_{M} = \{\theta_0, ..., \theta_m\} \sim D_{\Theta}$), to estimate a linear transformation matrix $A_{D}$. This matrix encodes the mapping between the sampling region $D_{\Theta}$ and a subspace $S_{D}$ within which the features are expected to vary most rapidly, and from where parameter samples should be drawn preferentially.

\begin{figure}[ht!]
  \centering
  \includegraphics[scale = 0.225]{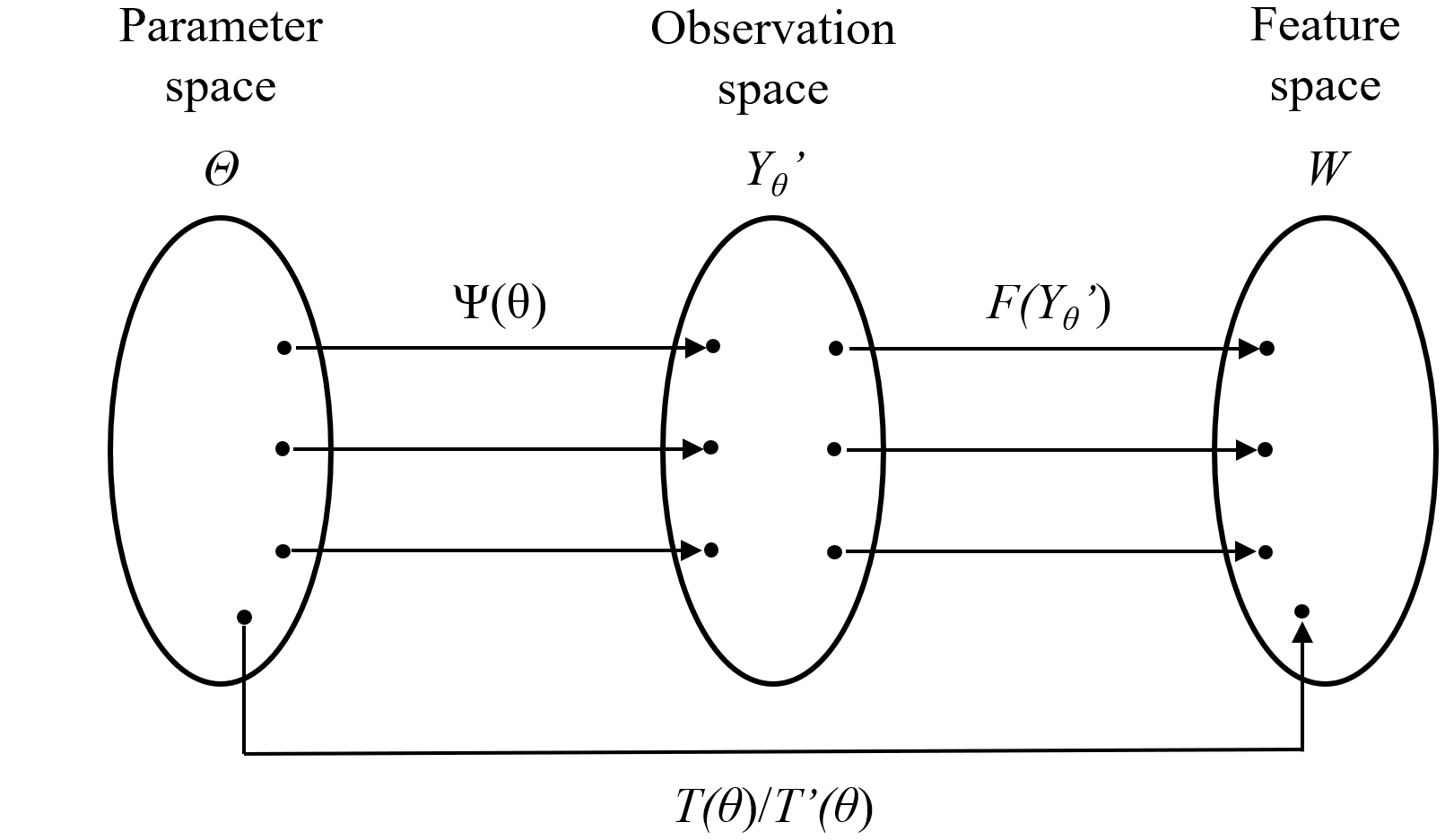}
  \caption{Diagram of the subspaces and mapping functions used in DR-FFIT.}
  \label{fig:fig1}
\end{figure}
The transformation matrices $A_{D}$ are derived by estimating an active subspace for $T'$ within $D_{\Theta}$. This is done by concatenating the $k$ largest loading eigenvectors of the second moment matrix $G$ of the estimated gradients $\nabla T'$: $G_{\theta_M} = E[\nabla T'(\theta_M)^T \nabla T'(\theta_M)]$ (Figure 2). In essence, $S_D$ represents the $k-$dimensional subspace that best captures the variance of the gradients of $T'$ within $D_{\Theta}$. 

\begin{figure}[ht!]
  \centering
  \includegraphics[scale = 0.3]{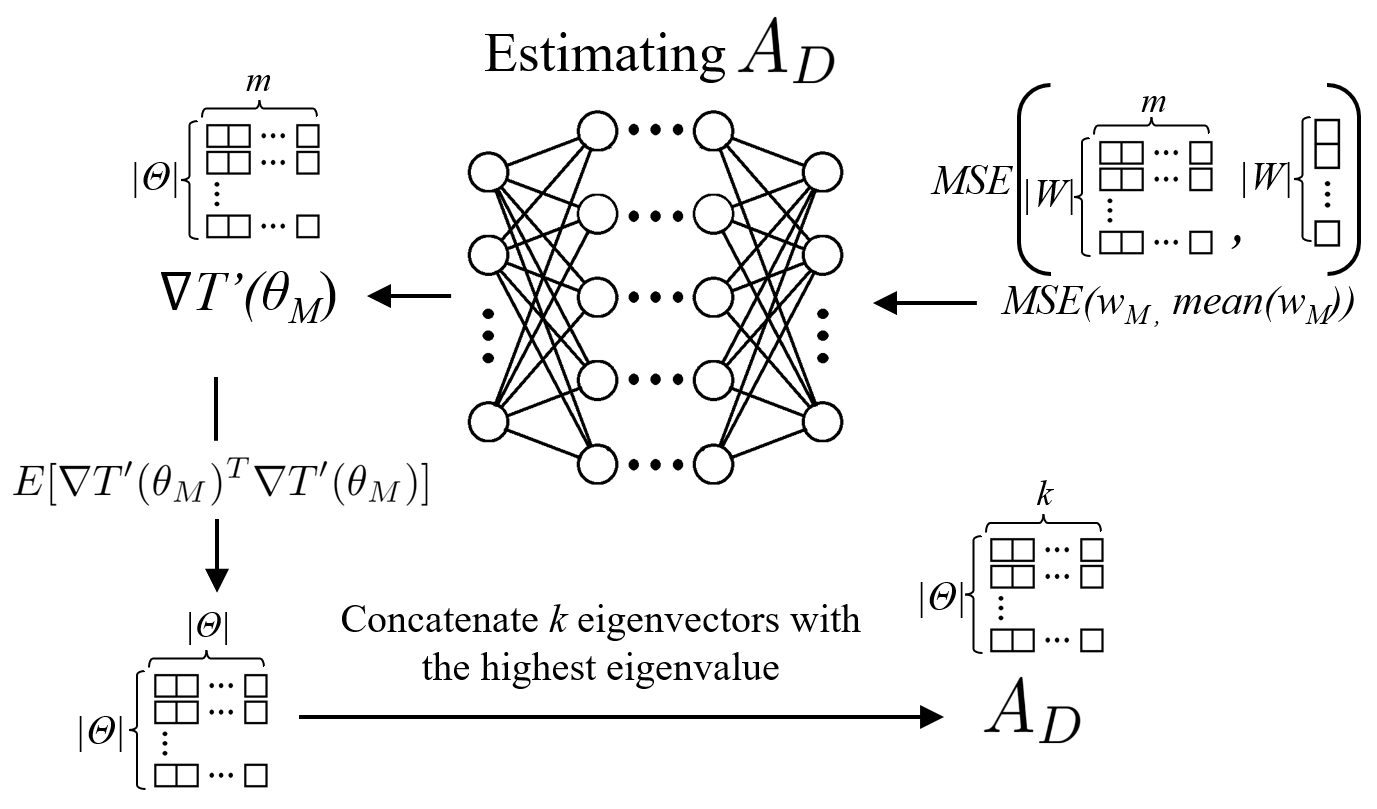}
  \caption{Estimation of transformation matrix $A_D$ via subspace estimation from a trained artificial neural network $T'$. The predicted feature values $w_M$ for samples $\theta_M$ (top right) are used to compute $\nabla T'(\theta_M)$ (top left), which is then used to estimate the second moment matrix $G$ (bottom left), which $k$ largest loading eigenvectors are used to obtain the transformation matrix $A_D$ (bottom right).}
  \label{fig:fig2}
\end{figure}

Let us consider $F$, $m$ and $k$ as hyperparameters of DR-FFIT, with $F$ and $k$ being the most sensitive for performance and $m$ being the most sensitive for run time. 

\section{Experiments}
We first describe the use-cases explored in the present report. We then explain the testing procedures between a selection of competing algorithms. Finally, we report the details specifying the search task performed by the algorithms.

\subsection{Use-cases}
We tested DR-FFIT on i) the Wilson-Cowan (WC) neural mass model with empirical magnetoencephalography (MEG) data and ii) a reduced Hodgkin-Huxley (HH) current clamp neuron model with synthetic data \cite{wilson_mathematical_1973, pospischil_minimal_2008}. All code is available on the \href{https://github.com/NeuroLife77/DRFFIT_paper}{github repository}. 

The WC model is designed to account for the dynamics of two coupled neural populations with recurrent connections, based on the following equations \cite{wilson_mathematical_1973,sanz-leon_mathematical_2015}:
\begin{equation}
\tau_e\frac{dE}{dt}=-E + (k_e - r_eE)S_e(c_{ee}E - c_{ie}I + P - \theta_e) 
\end{equation}
\begin{equation}
\tau_i\frac{dI}{dt}=-I + (k_i - r_iI)S_i(c_{ei}E - c_{ii}I + Q - \theta_i),
\end{equation}
where $E,I$ are state variables and $S_j(z)$ is a sigmoid activation function for a neural population $j \in \{e,i\}$:
\begin{equation}
S_j(z) = \frac{c_j}{1+exp[-a_j(\alpha_jz-b_j)]}.
\end{equation}
All the other terms are the parameters of the model (detailed in Table 1). We used two additional noise parameters ($\sigma_e,\sigma_i$), which account for the respective standard deviations of the noise affecting each population through the stochastic integration scheme. 

We define the WC numerical simulator $\Psi_{WC}(\theta) = Y_{\theta}'$ such that $ Y_{\theta}'$ is the power spectrum density (PSD) of the model output time series. We derived the PSDs using Welch's method, as implemented in SciPy \cite{virtanen_scipy_2020}, and evaluated the alignment between the model outputs and the empirical observations within the $2$-$80$Hz frequency range. The empirical observations were MEG resting-state recordings (i.e., task-free) from human participants, available from the Open MEG Archives (OMEGA) \cite{niso_omega_2016}, in the pre-processed version produced by \cite{da_silva_castanheira_brief_2021}. We used the negative Pearson correlation coefficient ($-R$) as loss function $\ell(Y,Y')$, and report $R$ as goodness of fit for simplicity.
\newpage
The other tested model is the HH model, which describes the state of a neuron under current clamp in terms of its membrane potential (see Table 2 for details on parameters) \cite{hodgkin_quantitative_1952}. We implemented a jit-complied version of the numerical simulator from the SBI toolkit's example section \cite{tejero-cantero_sbi_2020, lam_numba_2015}. The dynamics are described by the following equations: \cite{pospischil_minimal_2008}:
\begin{equation}
   C_m\frac{dV}{dt}=g_1\left(E_1-V\right)+
                    {\bar{g}_{Na}}m^3h\left(E_{Na}-V\right)+
                    {\bar{g}_{K}}n^4\left(E_K-V\right)+
                    \bar{g}_Mp\left(E_K-V\right)+
                    I_{inj}+
                    \sigma\eta\left(t\right),\\
\end{equation}
\begin{equation}
\frac{dq}{dt}=\frac{q_\infty\left(V\right)-q}{\tau_q\left(V\right)},\;q\in\{m,h,n,p\}.
\end{equation}
The state variable of this system is the membrane potential of the neuron, $V$; see Pospischil et al. 2008 for full details of the model and simulator \cite{pospischil_minimal_2008}. 

The numerical simulator of HH is defined as $\Psi_{HH}(\theta) = Y_{\theta}'$, such that $Y_{\theta}'$ is a time series of membrane potential. We used synthetic observations for this model and the mean-squared-error as $\ell(Y,Y')$ (all related plots herein report root-mean-squared-error values; RMSE). We produced the set of synthetic observations from 25,000 simulations sampled randomly over the entire parameter space, grouped into 25 clusters, obtained using the scikit-learn implementation of k-means \cite{pedregosa_scikit-learn_2011}. For each cluster, we identified the 10 closest and 10 farthest samples from each centroid, which we combined to create a set of 500 observations.

For both use-cases, we reduced their respective sets of empirical observations down to 20, keeping those for which initial solutions were associated with a consistently higher loss value for multiple runs. The initial solutions were obtained using the initialization step described below. After this selection, the same set of 20 empirical observations were used for each use-case across all the experiments reported.

\subsection{Comparison between algorithms}
We designed a two-step procedure consisting of an initialization step and of a search task to evaluate the tested algorithms. The initialization step for each observation was identical across the tested algorithms. We performed the complete experimental procedure twice, using initial values and training data for the NSEs, resulting in 2 trials per use-case, which we combined by looking at the improvement on initial solutions.

The initialization step produced both an initial solution for each observation and a training set for the NSEs to learn $T'$, as explained below. For each model and trial, we sampled $\epsilon = 10,000$ parameter candidates $\theta_{\epsilon}$ from a uniform distribution over the parameter space, computed the corresponding model predictions $Y_{\epsilon}'$ along with the respective loss $\ell(Y_i,Y_{\epsilon}')$ for all observations, and kept the respective best predictions $Y_{\theta}'$. For each trial, we used the initialization set containing all ($\theta_{\epsilon}$, $Y_{\epsilon}'$) pairs to derive the training set of ($\theta$, $w_{\theta}'$) pairs to train the NSEs. This process was completed identically across trials and for both use-cases.

We produced four sub-versions of DR-FFIT: two for RS and two for SA, each using either PCA or an auto-encoder (AE) as the feature function $F$. We fixed the feature space dimensionality to $|W| = 5$. We used  pytorch for all neural network implementations; we used scikit-learn to derives PCAs \cite{pedregosa_scikit-learn_2011,paszke_pytorch_2019}. The neural networks in NSEs consisted of stacks of single-layer residual blocks, with the SiLU activation function and batch normalization (see Table 3 for details). We trained the NSE networks using AdamW, with a step learning rate scheduler and L2-regularization with gradient clipping.  

\subsection{Search task}
We evaluated the convergence (based on the total number of samples used) and run time of each tested search method. We repeated each test 25 times for all 20 observations. For each model, we performed 100 iterations with a batch size of 25 samples and recorded the loss and run time. We used the same batch size and number of iterations for all tested algorithms to highlight the effect of their respective parameter-search method. We fixed the number of iterations instead of using a convergence threshold because our focus was on improvement in sampling efficiency rather than optimal loss. We also defined the mean relative difference (MRD) metric to evaluate the relative improvement on pure random search. We refer to the MRD value for the final solution after completion of the search task as final MRD (fMRD).

Because we compared the different search methods on equal grounds, the initialization step was applied for all tests that include a comparison with the DR-FFIT versions. To ensure that this did not hinder the performance of the other search methods we tested if they performed better by leveraging those samples in the search task instead of the initialization step. We maintained the same total number of samples, but instead of using 10,000 samples for the initialization step, we used 500 additional samples per observation during the search task.

\newpage
\section{Results}
\subsection{Use-case 1: Wilson-Cowan model}
We found that random-search augmented by DR-FFIT outperformed baseline random search: fMRD for  RS + AE: 10.37\%;  6.32\% for RS + PCA (Figure 3). We also observed substantial improvements for the DR-FFIT-augmented versions of simulated-annealing over RS: fMRD for SA + AE: 46.82\% and 41.15\% for SA + PCA, outperforming baseline SA (fMRD: 36.39\%). These results are shown Figure 3. 
Our data also show that DR-FFIT improved sampling efficiency across all implementations. DR-FFIT-augmented simulated-annealing with auto-encoder as feature function (SA + AE) outperformed the well-established CMA-ES both in terms of final loss value and convergence (Figure 3).

\begin{figure}[ht!]
  \centering
  \includegraphics[scale = 0.3]{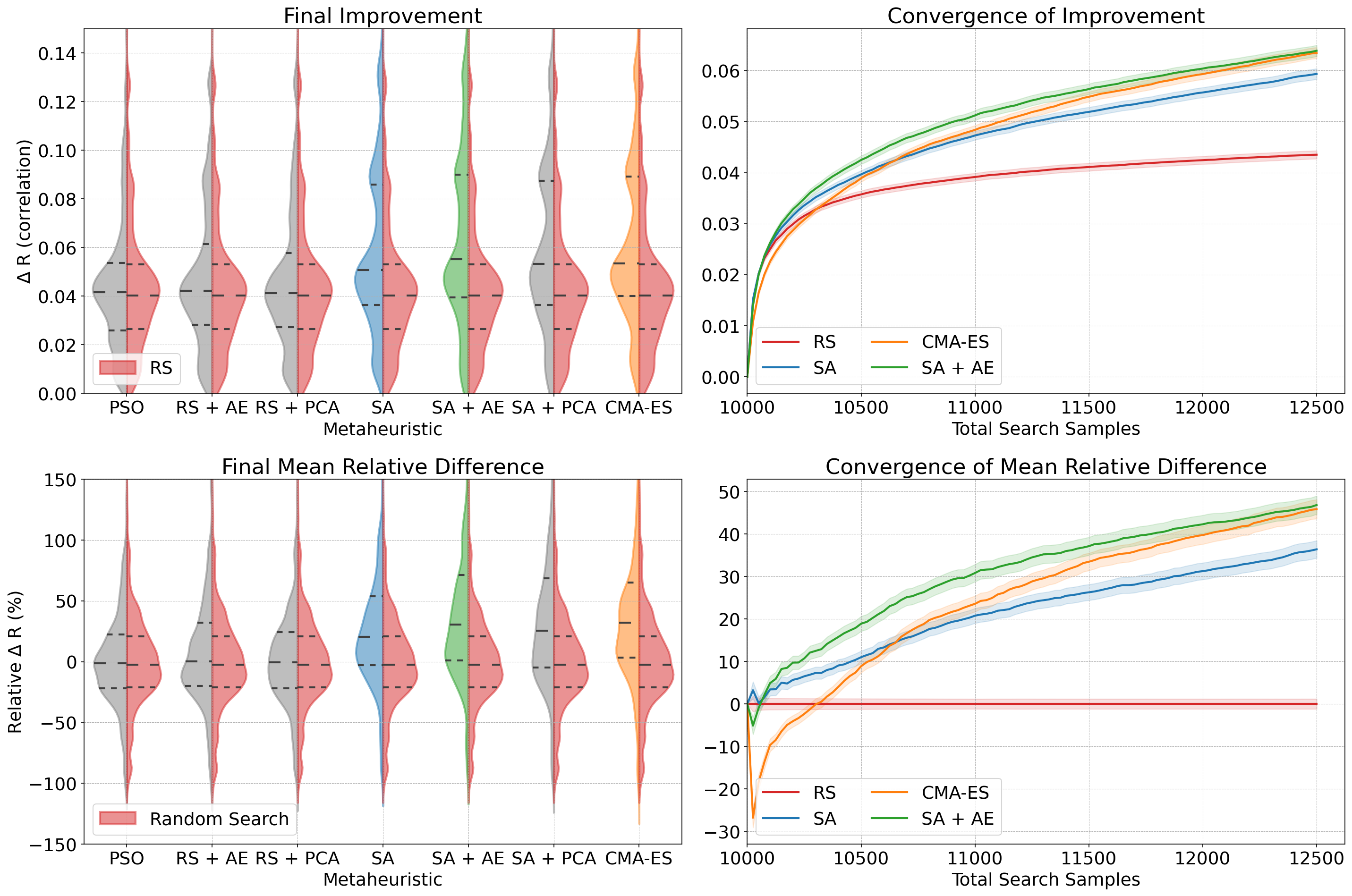}
  \caption{Performance improvements of DR-FFIT-augmented search methods (Wilson-Cowan use-case). Top row: (left) Empirical distributions of the performances obtained at the last iteration of the tested search methods ;  (right) the related average performances (shaded areas show standard-errors) as a function of the number of samples used per observation for 4 selected methods (right). Bottom row: (left) Empirical distributions of the fMRD obtained at the last iteration of the tested search methods; (right) average MRD (shaded areas show standard-errors) as a function of the number of samples used for 4 selected methods.}
  \label{fig:fig3}
\end{figure}

\begin{figure}[ht!]
  \centering
  \includegraphics[scale = 0.25]{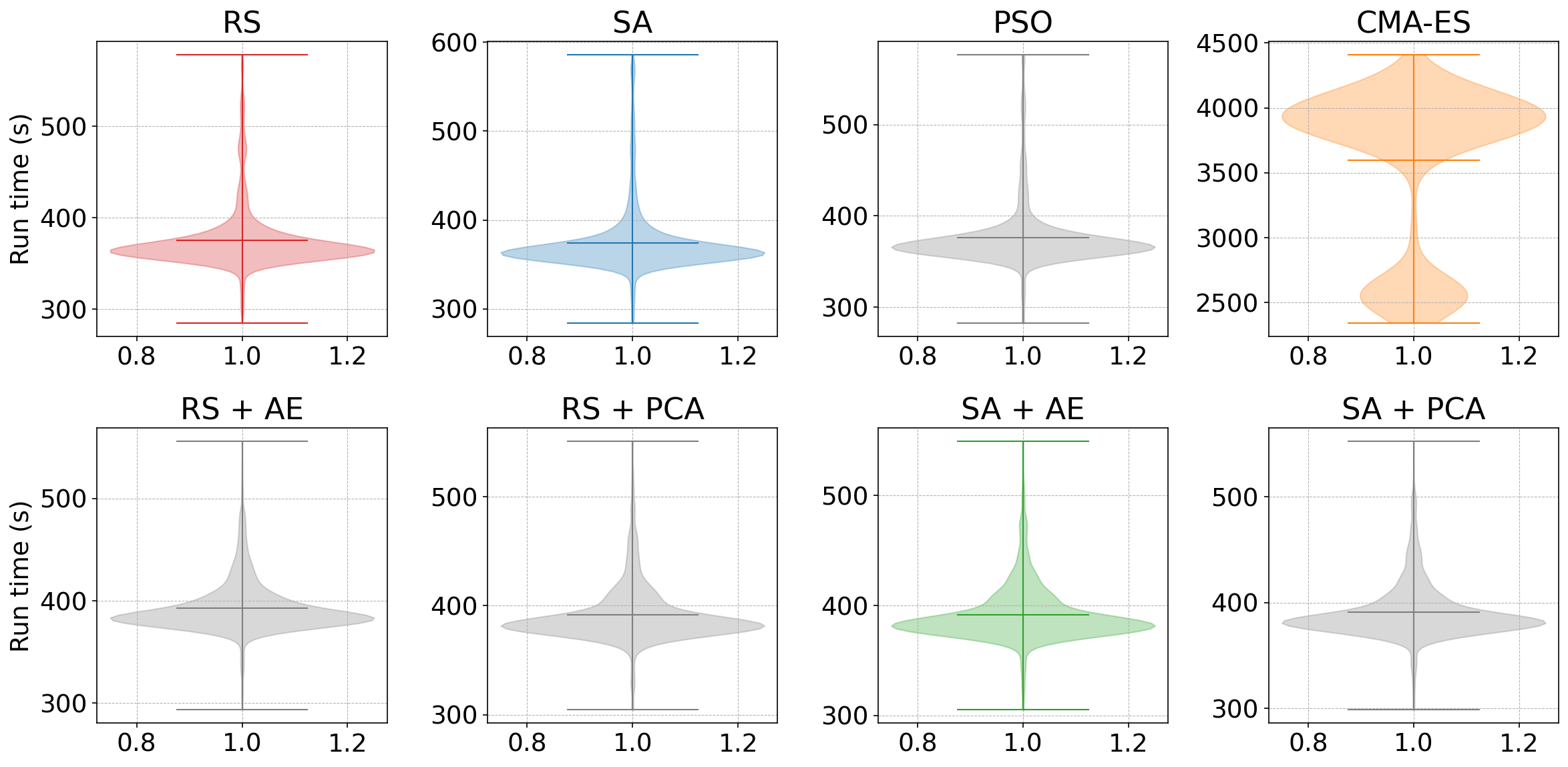}
  \caption{Empirical distribution of run-times of the tested search methods (Wilson-Cowan use-case).}
  \label{fig:fig4}
\end{figure}

The cost of CPU computing of $A_{D}$ increased the run time of all DR-FFIT-augmented algorithms by 16.9 seconds on average (a $<5\%$ relative increase) compared to the original versions (Figure 4). The CMA-ES metaheuristic's run-time was an order of magnitude longer than the other search methods', with an average of about 3,600 seconds (a 860\% relative increase with respect to RS). Overall, DR-FFIT did not affect run-time substantially, although $A_{D}$ was computed on the CPU, which is suboptimal. Our data show that, in the present experiment, the most efficient metaheurisitic was SA + AE: the DR-FFIT-augmented simulated annealing with the AE.

\subsection{Use-case 2: Hodgkin-Huxley model}
The DR-FFIT-augmented random searches (RS + AE and RS + PCA) outperformed baseline RS with a fMRD of 14.19\% for the AE version and 13.47\% for the PCA version (Figure 5). The DR-FFIT-augmented simulated annealing implementations (SA + AE and SA + PCA) outperformed baseline RS with a fMRD of 17.21\% for the AE version and 21.82\% for the PCA version. They also both outperformed baseline SA, which showcased an fMRD of 4.64\% (Figure 5). The convergence results demonstrate that the DR-FFIT metaheuristic improved sampling efficiency across all implementations.

\begin{figure}[ht!]
  \centering
  \includegraphics[scale = 0.3]{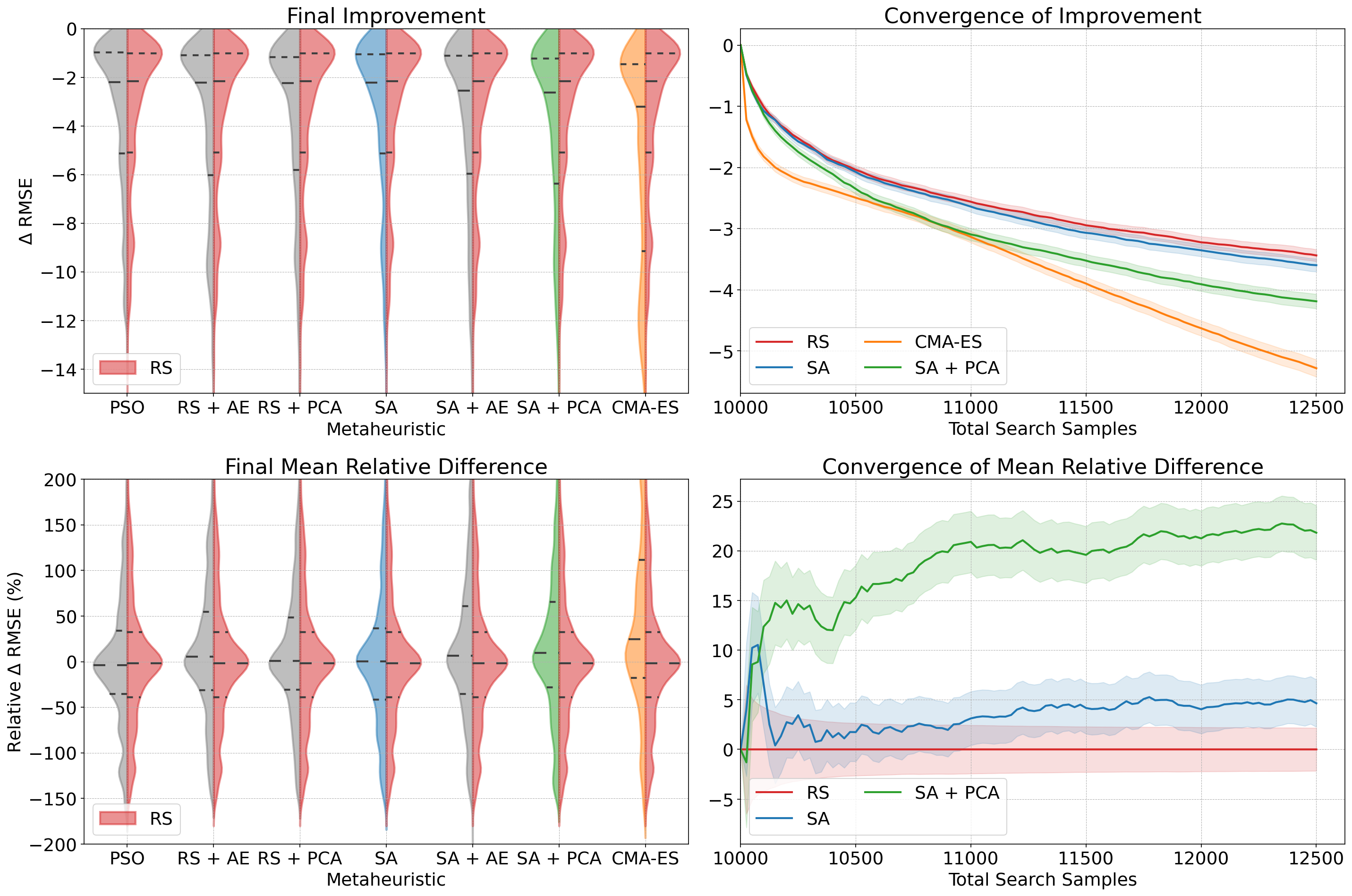}
  \caption{Performance improvements of DR-FFIT-augmented search methods (Hodgkin-Huxley use-case). Top row: (left) Empirical distributions of the performances obtained at the last iteration of the tested search methods ;  (right) the related average performances (shaded areas show standard-errors) as a function of the number of samples used per observation for 4 selected methods (right). Bottom row: (left) Empirical distributions of the fMRD obtained at the last iteration of the tested search methods; (right) average MRD (shaded areas show standard-errors) as a function of the number of samples used for 3 of the 4 selected methods.}
  \label{fig:fig5}
\end{figure}

The cost of computing $A_{D}$ on the CPU increased the run-time of DR-FFIT-augmented algorithms by 21.1 seconds on average (a 60.5\% relative increase with respect to baseline algorithms), compared to the original RS and SA methods (Figure 6). The longer run-time for the AE versions is due to the larger size of their respective neural networks (see Table 3). CMA-ES featured a run-time an order of magnitude longer than the other search methods', with an average of 388.07 seconds (a 981\% increase with respect to RS). Thus, the increase in run-time induced by DR-FFIT remains considerably smaller than with CMA-ES, the only metaheuristic that yielded a greater improvement over the initial solution than DR-FFIT in this use-case, albeit with a considerably longer run-time.

\begin{figure}[ht!]
  \centering
  \includegraphics[scale = 0.25]{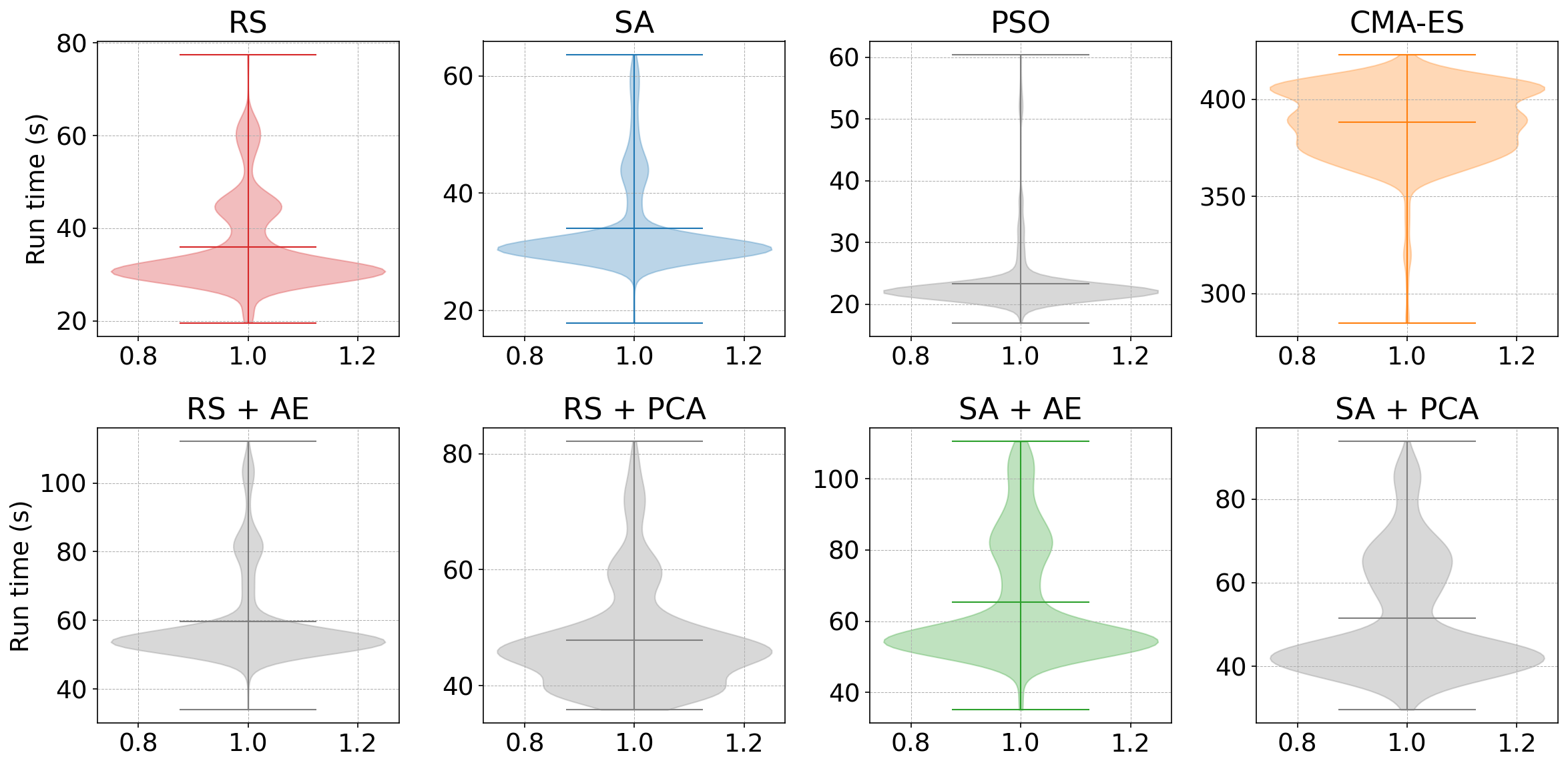}
  \caption{Distribution of run-times of the tested search methods (Hodgkin-Huxley use-case).}
  \label{fig:fig6}
\end{figure}
Overall, our data show that the DR-FFIT metaheuristic is computationally affordable and increases performances without inducing substantial increases in run-time, compared to alternatives such as CMA-ES, even in suboptimal conditions of execution (i.e., on CPU).

\subsection{Impact of initialization}
In both use-cases, all search methods resulted in a significantly higher loss when implemented without the initialization step (Figure 7 and 8). In fact, none of the tested search methods yielded solutions with a lower associated loss than with the initialization step alone. In the first use-case, on average, the best solution across runs without initialization step has a pearson correlation goodness of fit value of 0.8004 versus 0.8231 for the solutions from the initialization step alone. In the second use-case, on average, the RMSE loss was of 33.53 for the runs without initialization step versus a lower RMSE loss 30.21 for the solutions from the initialization step alone.  

\begin{figure}[ht!]
  \centering
  \includegraphics[scale = 0.3]{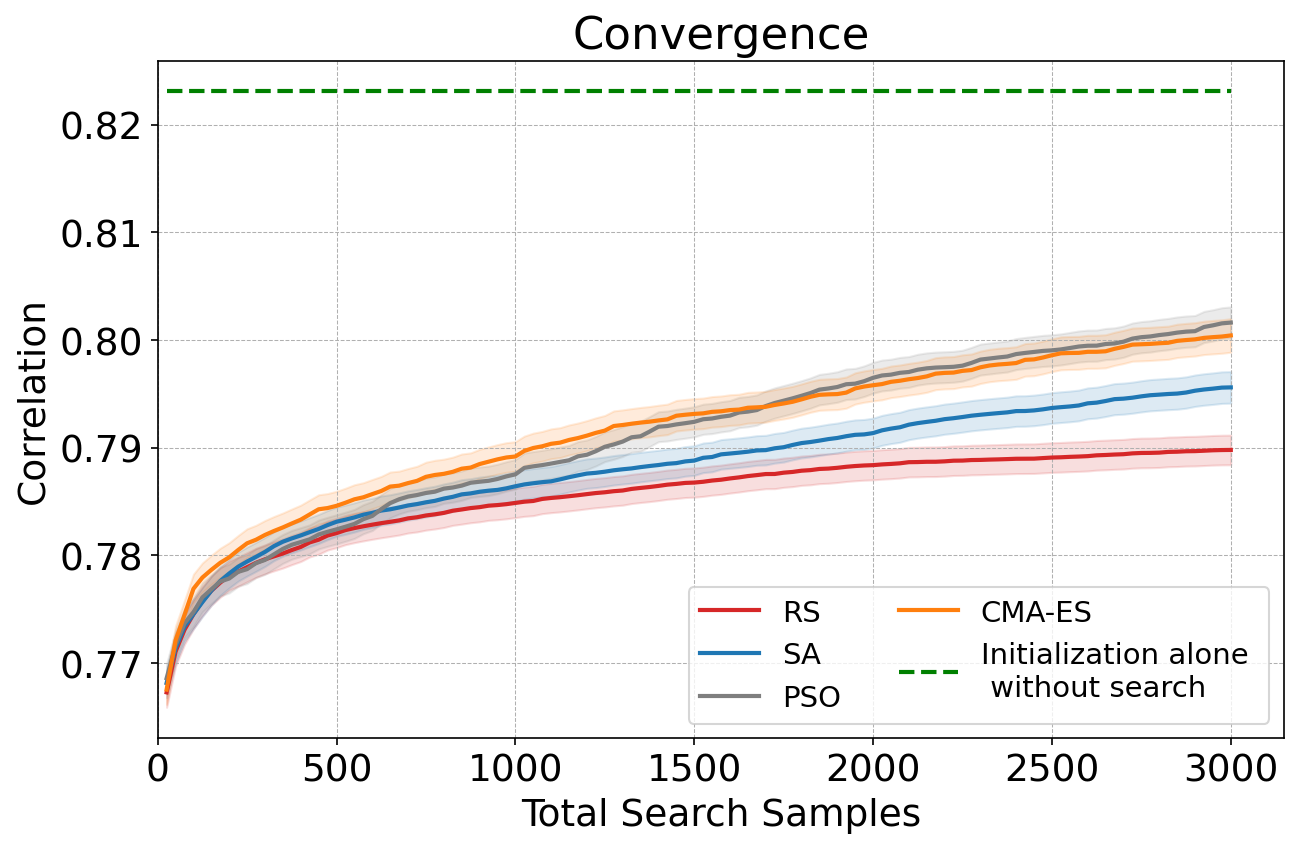}
  \hspace{0.25cm}
  \includegraphics[scale = 0.2]{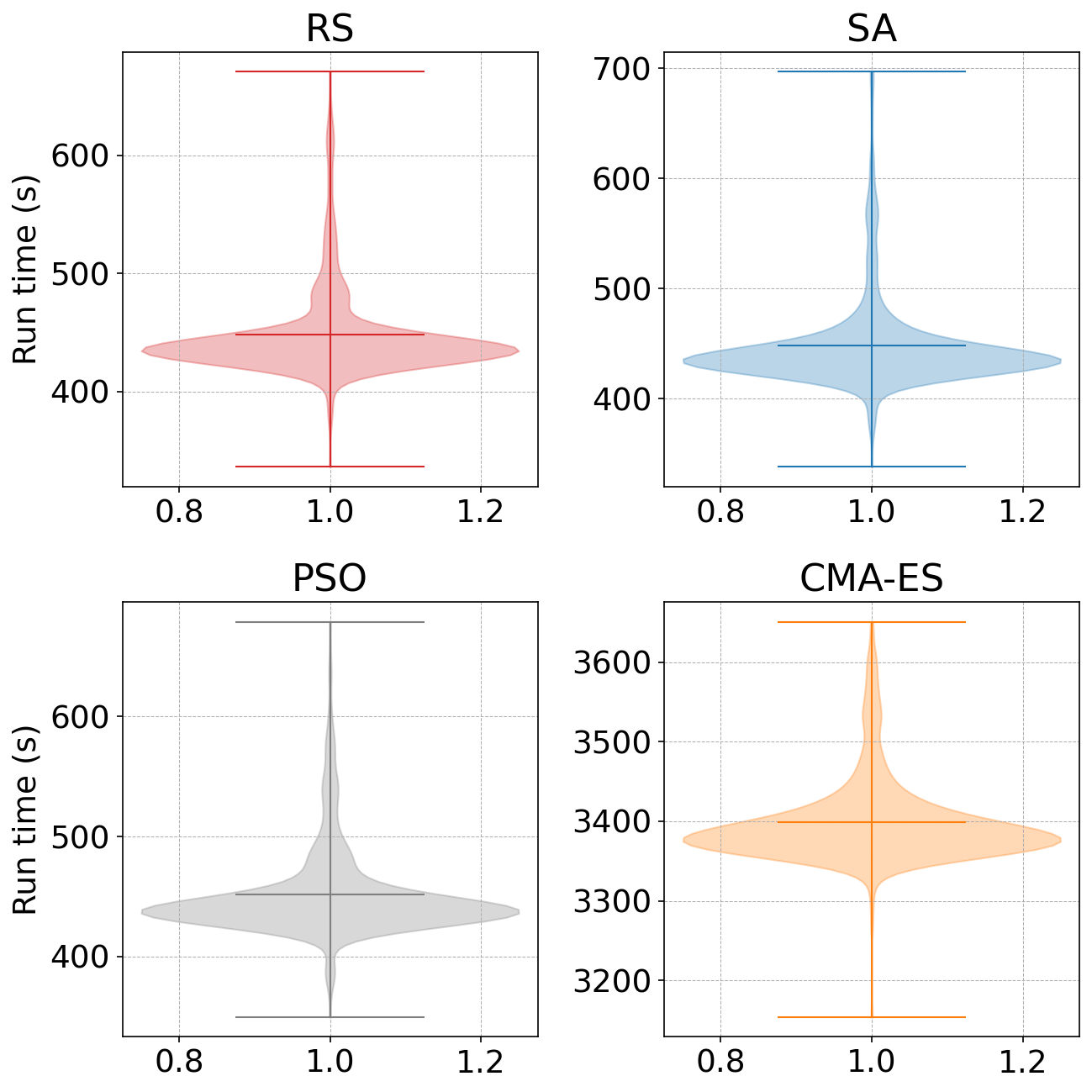}
  \caption{Left: Convergence shown as average goodness-of-fit (with standard error) with respect to the number of samples used in the search. Right: empirical distributions of run-time, in seconds, for the benchmark search methods tested (Wilson-Cowan use-case).}
  \label{fig:fig7}
\end{figure}

\begin{figure}[ht!]
  \centering
  \includegraphics[scale = 0.3]{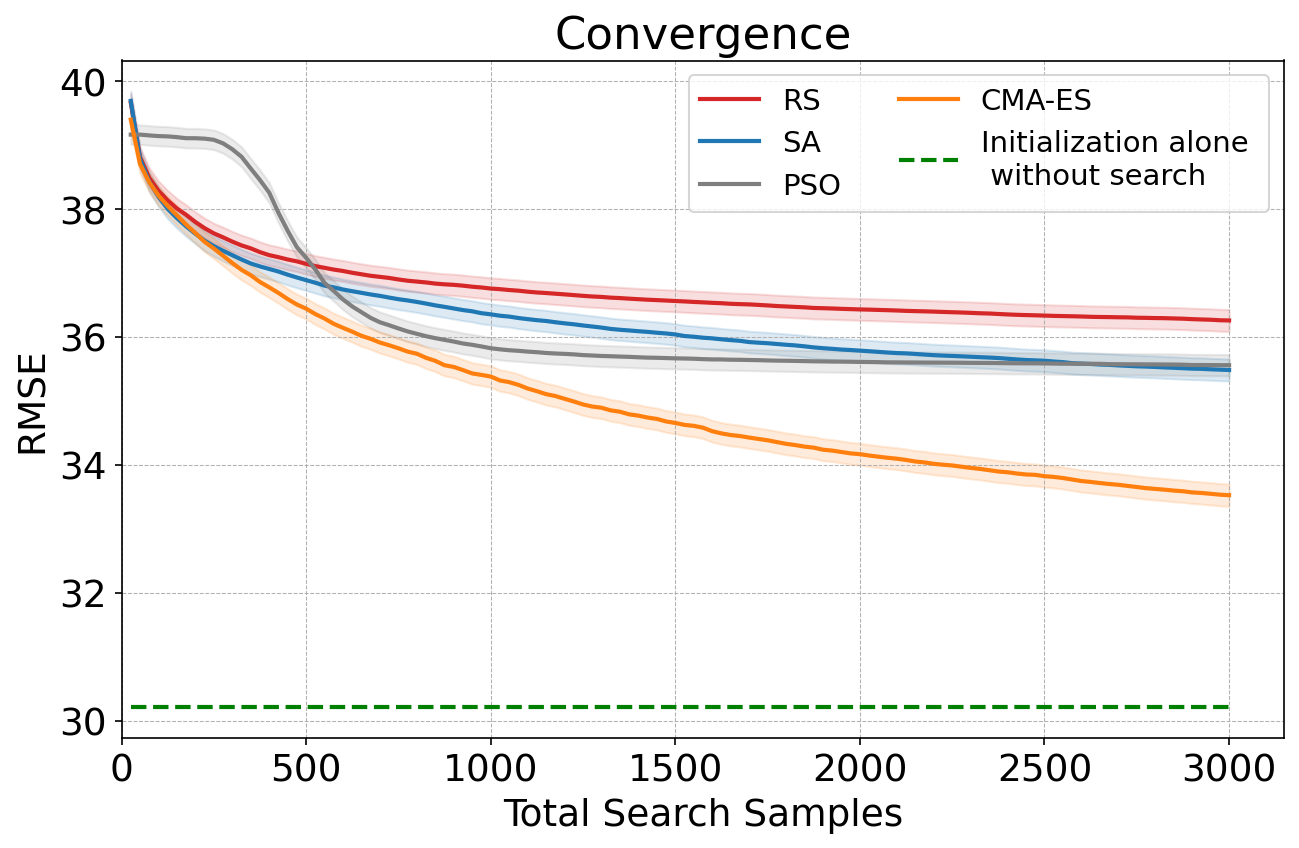}
  \hspace{0.25cm}
  \includegraphics[scale = 0.2]{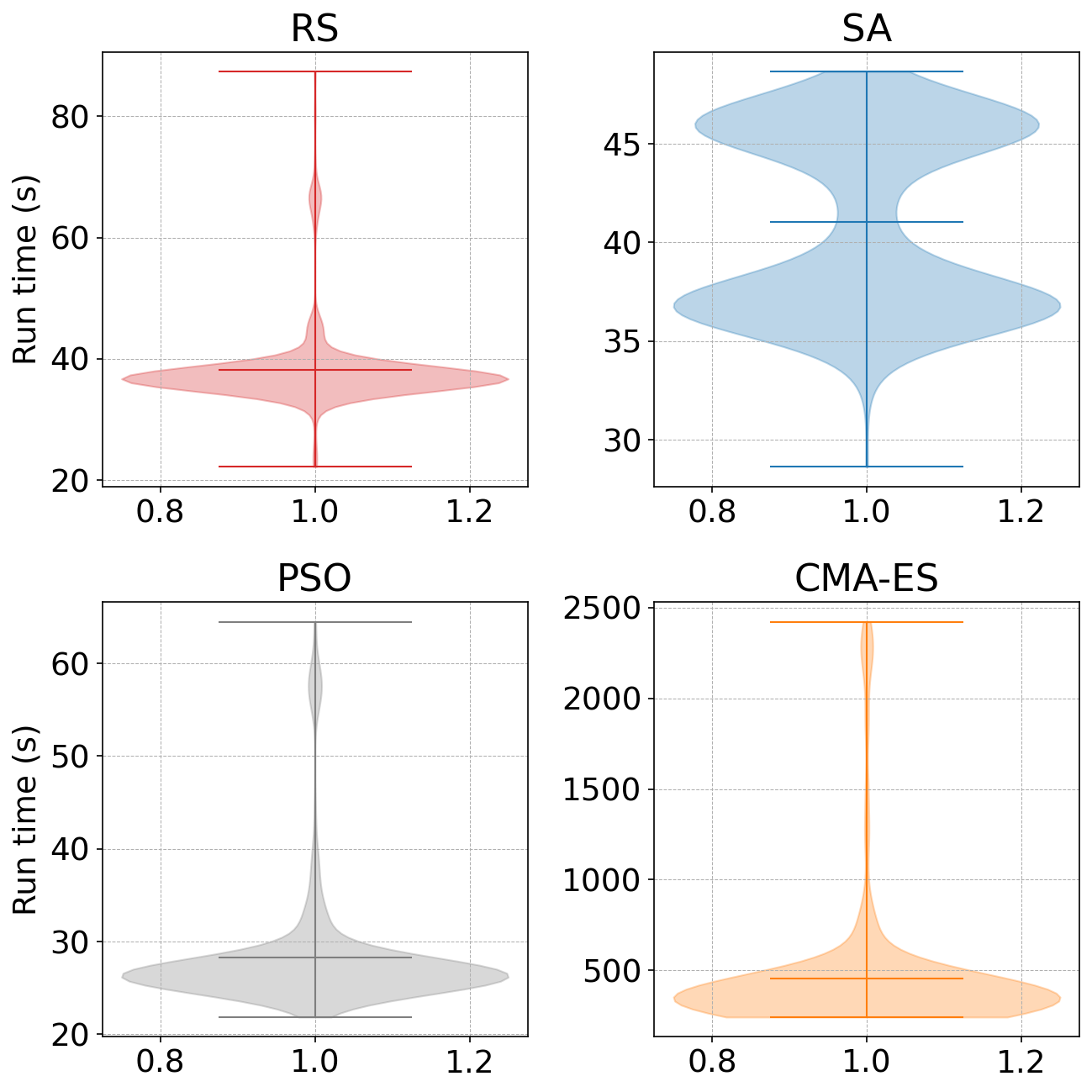}
  \caption{Left: Convergence shown as average loss (with standard error) with respect to the number of samples used in the search. Right: empirical distributions of run-time, in seconds, for the benchmark search methods tested (Hodgkin-Huxley use-case).}
  \label{fig:fig8}
\end{figure}

These results show that the relative performances of baseline vs. DR-FFIT-augmented search methods did not depend on the imposed initialization step.
\section{Conclusion}
We introduced DR-FFIT, a novel metaheuristic that improves sampling efficiency of search methods in large parameter spaces reduction. We tested DR-FFIT in two use-cases inspired by current challenges in computational neuroscience. Our data demonstrate that DR-FFIT improves the convergence and overall performances of search outcomes, with limited additional computational cost, and considerably lesser run time than competing alternatives. The initialization step imposed by DR-FFIT contributed to the improved performances of all tested algorithms performing parameter search for a set of 20 distinct observations with a given model. We conclude that DR-FFIT is a competitive metaheuristic that may prove useful tool in a variety of optimization challenges where point estimates of the free parameters of a model need to be derived for multiple observations.
\newpage
\section{Tables}
\begin{table}[ht!]
		\centering
		\begin{tabular}{|c | c |}
			\hline
			Parameter & Description \\ \hline
			$c_{ee}$ & Excitatory to excitatory coupling coefficient\\ \hline
			$c_{ei}$ & Excitatory to inhibitory coupling coefficient\\ \hline
			$c_{ie}$ & Inhibitory to excitatory coupling coefficient\\ \hline
			$c_{ii}$ & Inhibitory to inhibitory coupling coefficient\\ \hline
			$\tau_e$ & Excitatory population, membrane time-constant\\ \hline
			$\tau_i$ & Inhibitory population, membrane time-constant\\ \hline
			$a_e$ & Value of the maximum slope of the sigmoid function $S_e$\\ \hline
			$b_e$ & Sigmoid function threshold\\ \hline
			$a_i$ & Value of the maximum slope of the sigmoid function $S_i$\\ \hline
			$b_i$ & Sigmoid function threshold \\ \hline
			$\theta_e$ & Position of the maximum slope of $S_e$\\ \hline
			$\theta_i$ & Position of the maximum slope of $S_i$\\ \hline	
			$r_i$ & Inhibitory refractory period\\ \hline
			$r_e$ & Inhibitory refractory period\\ \hline
			$k_e$ & Maximum value of the excitatory response function\\ \hline	
			$k_i$ &	Maximum value of the inhibitory response function\\ \hline	
			$\alpha_e$ & Balance parameter between excitatory and inhibitory masses\\ \hline
			$\alpha_i$ & Balance parameter between excitatory and inhibitory masses\\ \hline			
			$P$ & External stimulus to the excitatory population\\ \hline
			$Q$ & External stimulus to the inhibitory population\\ \hline
			$\sigma_e$  & Variance of noise amplitude on state variable E\\ \hline
			$\sigma_i$  & Variance of noise amplitude on state variable I\\ \hline
		\end{tabular}
		\caption{Parameters for the Wilson-Cowan simulator}
		\label{table:1}
	\end{table}
	
\begin{table}[ht!]
		\centering
		\begin{tabular}{|c | c |}
			\hline
			Parameter & Description \\ \hline
			$C_m$ & Membrane capacitance\\ \hline
			$V_T$ & Adjustment variable for the spike threshold \\ \hline
			$g_{\text{l}}$ & Leak conductance\\ \hline
			$E_{\text{l}}$ & Membrane reversal potential\\ \hline
			$\bar{g}_K$ & Density of potassium (K) channels\\ \hline
			$\bar{g}_{Na}$ & Density of sodium (Na) channels\\ \hline
			$\bar{g}_M$ & Density of a slow slow voltage-dependent potassium (K) channel\\ \hline
			$E_K$ & Reversal potential of potassium (K) ions\\ \hline
			$E_{Na}$ & Reversal potential of sodium (Na) ions\\ \hline
			$\tau_{max}$ & Maximum value of the time constant\\ \hline
			$\sigma$ & Intrinsic neural noise\\ \hline
			$I$ & Injected current\\ \hline
		\end{tabular}
		\caption{Parameters for the Hodgkin-Huxley simulator}
		\label{table:2}
	\end{table}

\begin{table}[ht!]
		\centering
		\begin{tabular}{| c | c | c | c | c | c |}
			\hline
			Model & Trial & Feature function & Architecture & Activation function & Skip connection\\ \hline
			\multirow{4}{*}{WC} & \multirow{2}{*}{Trial 1} & AE & [100, 50, 50] & SiLU & True\\ \cline{3-6}
			& & PCA & [100, 50, 50] & SiLU & True\\ \cline{2-6}
			&\multirow{2}{*}{Trial 2} & AE & [100, 50, 50, 50] & SiLU & True\\ \cline{3-6}
			&  & PCA & [100, 50, 50] & SiLU & True\\  \cline{1-6}
			  
			\multirow{4}{*}{HH} & \multirow{2}{*}{Trial 1} & AE & [50, x 8] & SiLU & True\\ \cline{3-6}
			& & PCA & [50, 50, 50, 50] & PReLU & False\\ \cline{2-6}
			& \multirow{2}{*}{Trial 2} & AE & [50, x 8] & SiLU & True\\ \cline{3-6}
			&  & PCA & [50, 50] & PReLU & False\\ 
			  
			  \hline
		\end{tabular}
		\caption{Architecture for the neural networks $T'$}
		\label{table:3}
	\end{table}

\newpage

\section*{Acknowledgments}
This work was made possible thanks to a NSERC Discovery grant, the CIHR Canada Research Chair (Tier-1) of Neural Dynamics of Brain Systems, and the National Institutes of Health (1R01EB026299).

\bibliographystyle{unsrt}  
\bibliography{fixed_references}  

\begin{thebibliography}{10}

\bibitem{cranmer_frontier_2020}
Kyle Cranmer, Johann Brehmer, and Gilles Louppe.
\newblock The frontier of simulation-based inference.
\newblock {\em Proceedings of the National Academy of Sciences},
  117(48):30055--30062, December 2020.
\newblock Publisher: Proceedings of the National Academy of Sciences.

\bibitem{villaverde_benchmarking_2019}
Alejandro~F. Villaverde, Fabian Fr{\"o}hlich, Daniel Weindl, Jan Hasenauer, and
  Julio~R. Banga.
\newblock Benchmarking optimization methods for parameter estimation in large
  kinetic models.
\newblock {\em Bioinformatics (Oxford, England)}, 35(5):830--838, March 2019.

\bibitem{schalte_evaluation_2018}
Yannik Sch{\"a}lte, Paul Stapor, and Jan Hasenauer.
\newblock Evaluation of {Derivative}-{Free} {Optimizers} for {Parameter}
  {Estimation} in {Systems} {Biology}.
\newblock {\em IFAC-PapersOnLine}, 51:98--101, January 2018.

\bibitem{deistler_energy-efficient_2022}
Michael Deistler, Jakob~H. Macke, and Pedro~J. Gon{\c c}alves.
\newblock Energy-efficient network activity from disparate circuit parameters.
\newblock {\em Proceedings of the National Academy of Sciences},
  119(44):e2207632119, November 2022.
\newblock Publisher: Proceedings of the National Academy of Sciences.

\bibitem{karnopp_random_1963}
Dean~C. Karnopp.
\newblock Random search techniques for optimization problems.
\newblock {\em Automatica}, 1(2):111--121, August 1963.

\bibitem{yang_chapter_2021-1}
Xin-She Yang.
\newblock Chapter 5 - {Simulated} {Annealing}.
\newblock In Xin-She Yang, editor, {\em Nature-{Inspired} {Optimization}
  {Algorithms} ({Second} {Edition})}, pages 83--90. Academic Press, January
  2021.

\bibitem{yang_chapter_2021}
Xin-She Yang.
\newblock Chapter 8 - {Particle} {Swarm} {Optimization}.
\newblock In Xin-She Yang, editor, {\em Nature-{Inspired} {Optimization}
  {Algorithms} ({Second} {Edition})}, pages 111--121. Academic Press, January
  2021.

\bibitem{jastrebski_improving_2006}
G.A. Jastrebski and D.V. Arnold.
\newblock Improving {Evolution} {Strategies} through {Active} {Covariance}
  {Matrix} {Adaptation}.
\newblock In {\em 2006 {IEEE} {International} {Conference} on {Evolutionary}
  {Computation}}, pages 2814--2821, July 2006.
\newblock ISSN: 1941-0026.

\bibitem{wilson_mathematical_1973}
H.~R. Wilson and J.~D. Cowan.
\newblock A mathematical theory of the functional dynamics of cortical and
  thalamic nervous tissue.
\newblock {\em Kybernetik}, 13(2):55--80, September 1973.

\bibitem{baillet_magnetoencephalography_2017}
Sylvain Baillet.
\newblock Magnetoencephalography for brain electrophysiology and imaging.
\newblock {\em Nature Neuroscience}, 20(3):327--339, March 2017.
\newblock Number: 3 Publisher: Nature Publishing Group.

\bibitem{niso_omega_2016}
Guiomar Niso, Christine Rogers, Jeremy~T. Moreau, Li-Yuan Chen, Cecile Madjar,
  Samir Das, Elizabeth Bock, Fran{\c c}ois Tadel, Alan~C. Evans, Pierre
  Jolicoeur, and Sylvain Baillet.
\newblock {OMEGA}: {The} {Open} {MEG} {Archive}.
\newblock {\em NeuroImage}, 124:1182--1187, January 2016.

\bibitem{hodgkin_quantitative_1952}
A.~L. Hodgkin and A.~F. Huxley.
\newblock A quantitative description of membrane current and its application to
  conduction and excitation in nerve.
\newblock {\em The Journal of Physiology}, 117(4):500--544, August 1952.

\bibitem{pospischil_minimal_2008}
Martin Pospischil, Maria Toledo-Rodriguez, Cyril Monier, Zuzanna Piwkowska,
  Thierry Bal, Yves Fr{\'e}gnac, Henry Markram, and Alain Destexhe.
\newblock Minimal {Hodgkin}-{Huxley} type models for different classes of
  cortical and thalamic neurons.
\newblock {\em Biological Cybernetics}, 99(4-5):427--441, November 2008.

\bibitem{hansen_cma-espycma_2023}
Nikolaus Hansen, Youhei Akimoto, and Petr Baudis.
\newblock {CMA-ES/pycma} on {G}ithub.
\newblock Zenodo, DOI:10.5281/zenodo.2559634, February 2019.

\bibitem{sanz-leon_mathematical_2015}
Paula Sanz-Leon, Stuart~A. Knock, Andreas Spiegler, and Viktor~K. Jirsa.
\newblock Mathematical framework for large-scale brain network modeling in
  {The} {Virtual} {Brain}.
\newblock {\em NeuroImage}, 111:385--430, May 2015.

\bibitem{virtanen_scipy_2020}
Pauli Virtanen, Ralf Gommers, Travis~E. Oliphant, Matt Haberland, Tyler Reddy,
  David Cournapeau, Evgeni Burovski, Pearu Peterson, Warren Weckesser, Jonathan
  Bright, St{\'e}fan~J. van~der Walt, Matthew Brett, Joshua Wilson, K.~Jarrod
  Millman, Nikolay Mayorov, Andrew R.~J. Nelson, Eric Jones, Robert Kern, Eric
  Larson, C.~J. Carey, {\.I}lhan Polat, Yu~Feng, Eric~W. Moore, Jake
  VanderPlas, Denis Laxalde, Josef Perktold, Robert Cimrman, Ian Henriksen,
  E.~A. Quintero, Charles~R. Harris, Anne~M. Archibald, Ant{\^o}nio~H. Ribeiro,
  Fabian Pedregosa, and Paul van Mulbregt.
\newblock {SciPy} 1.0: fundamental algorithms for scientific computing in
  {Python}.
\newblock {\em Nature Methods}, 17(3):261--272, March 2020.
\newblock Number: 3 Publisher: Nature Publishing Group.

\bibitem{da_silva_castanheira_brief_2021}
Jason da~Silva~Castanheira, Hector~Domingo Orozco~Perez, Bratislav Misic, and
  Sylvain Baillet.
\newblock Brief segments of neurophysiological activity enable individual
  differentiation.
\newblock {\em Nature Communications}, 12(1):5713, September 2021.

\bibitem{tejero-cantero_sbi_2020}
Alvaro Tejero-Cantero, Jan Boelts, Michael Deistler, Jan-Matthis Lueckmann,
  Conor Durkan, Pedro~J. Gon{\c c}alves, David~S. Greenberg, and Jakob~H.
  Macke.
\newblock sbi: {A} toolkit for simulation-based inference.
\newblock {\em Journal of Open Source Software}, 5(52):2505, August 2020.

\bibitem{lam_numba_2015}
Siu~Kwan Lam, Antoine Pitrou, and Stanley Seibert.
\newblock Numba: a {LLVM}-based {Python} {JIT} compiler.
\newblock In {\em Proceedings of the {Second} {Workshop} on the {LLVM}
  {Compiler} {Infrastructure} in {HPC}}, {LLVM} '15, pages 1--6, New York, NY,
  USA, November 2015. Association for Computing Machinery.

\bibitem{pedregosa_scikit-learn_2011}
Fabian Pedregosa, Ga{\"e}l Varoquaux, Alexandre Gramfort, Vincent Michel,
  Bertrand Thirion, Olivier Grisel, Mathieu Blondel, Peter Prettenhofer, Ron
  Weiss, Vincent Dubourg, Jake Vanderplas, Alexandre Passos, David Cournapeau,
  Matthieu Brucher, Matthieu Perrot, and {\'E}douard Duchesnay.
\newblock Scikit-learn: {Machine} {Learning} in {Python}.
\newblock {\em The Journal of Machine Learning Research}, 12(null):2825--2830,
  November 2011.

\bibitem{paszke_pytorch_2019}
Adam Paszke, Sam Gross, Francisco Massa, Adam Lerer, James Bradbury, Gregory
  Chanan, Trevor Killeen, Zeming Lin, Natalia Gimelshein, Luca Antiga, Alban
  Desmaison, Andreas Kopf, Edward Yang, Zachary DeVito, Martin Raison, Alykhan
  Tejani, Sasank Chilamkurthy, Benoit Steiner, Lu~Fang, Junjie Bai, and Soumith
  Chintala.
\newblock {PyTorch}: {An} {Imperative} {Style}, {High}-{Performance} {Deep}
  {Learning} {Library}.
\newblock In {\em Advances in {Neural} {Information} {Processing} {Systems}},
  volume~32. Curran Associates, Inc., 2019.

\end{thebibliography}
\newpage
\section{Supplementary}
\subsection{Figures}
\begin{suppfigure}[ht!]
  \centering
  \includegraphics[scale = 0.27]{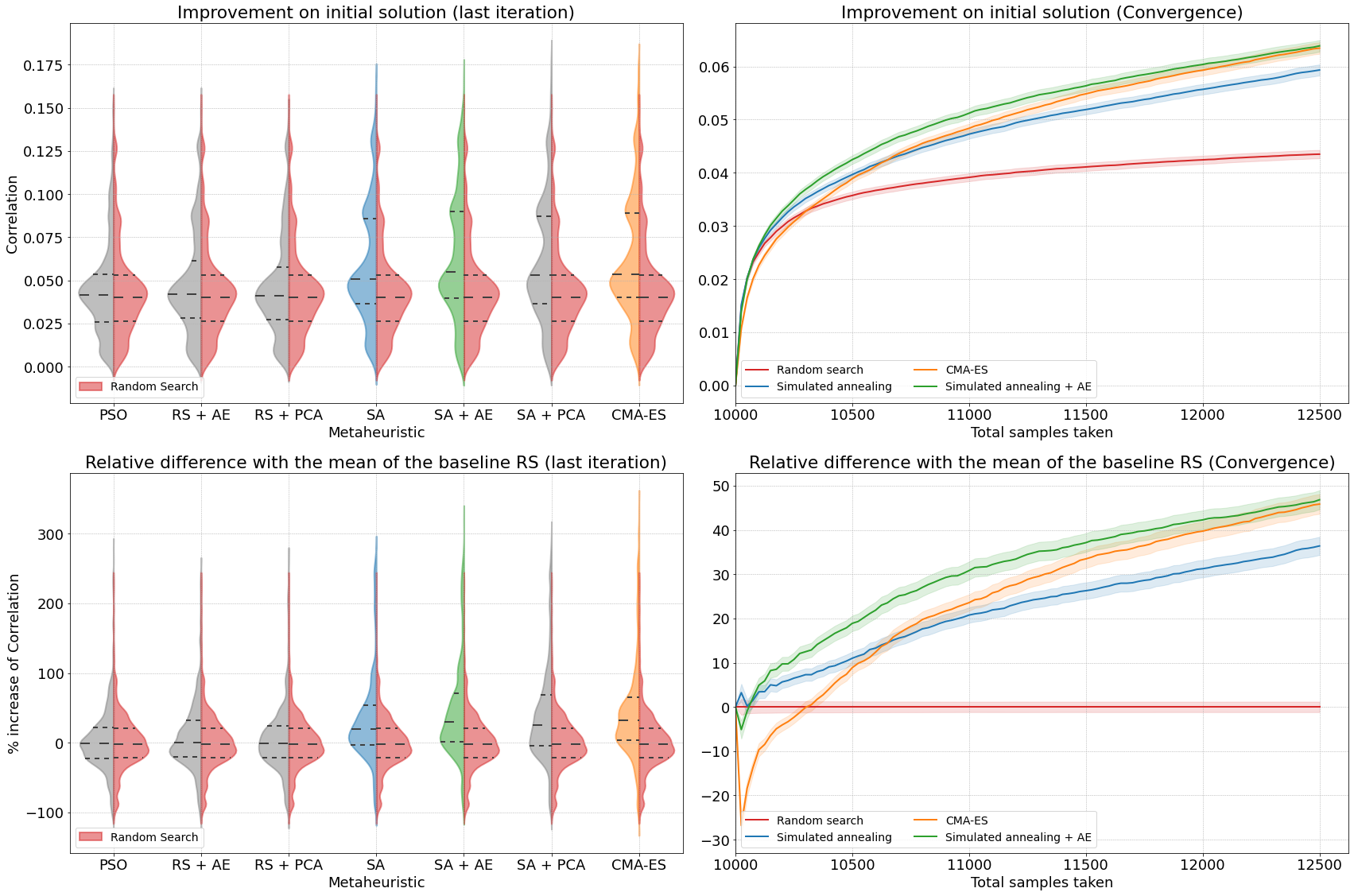}
  \caption{Performance of DR-FFIT-augmented search methods (Wilson-Cowan use-case). Same data as for Figure 3, with full empirical  distributions (left plots). Top row: (left) Empirical distributions of the performances obtained at the last iteration of the tested search methods ;  (right) the related average performances (shaded areas show standard-errors) as a function of the number of samples used per observation for 4 selected methods (right). Bottom row: (left) Empirical distributions of the fMRD obtained at the last iteration of the tested search methods; (right) average MRD (shaded areas show standard-errors) as a function of the number of samples used for the 4 selected methods.}
  \label{suppfigure}
\end{suppfigure}

\newpage
\begin{suppfigure}[ht!]
  \centering
  \includegraphics[scale = 0.27]{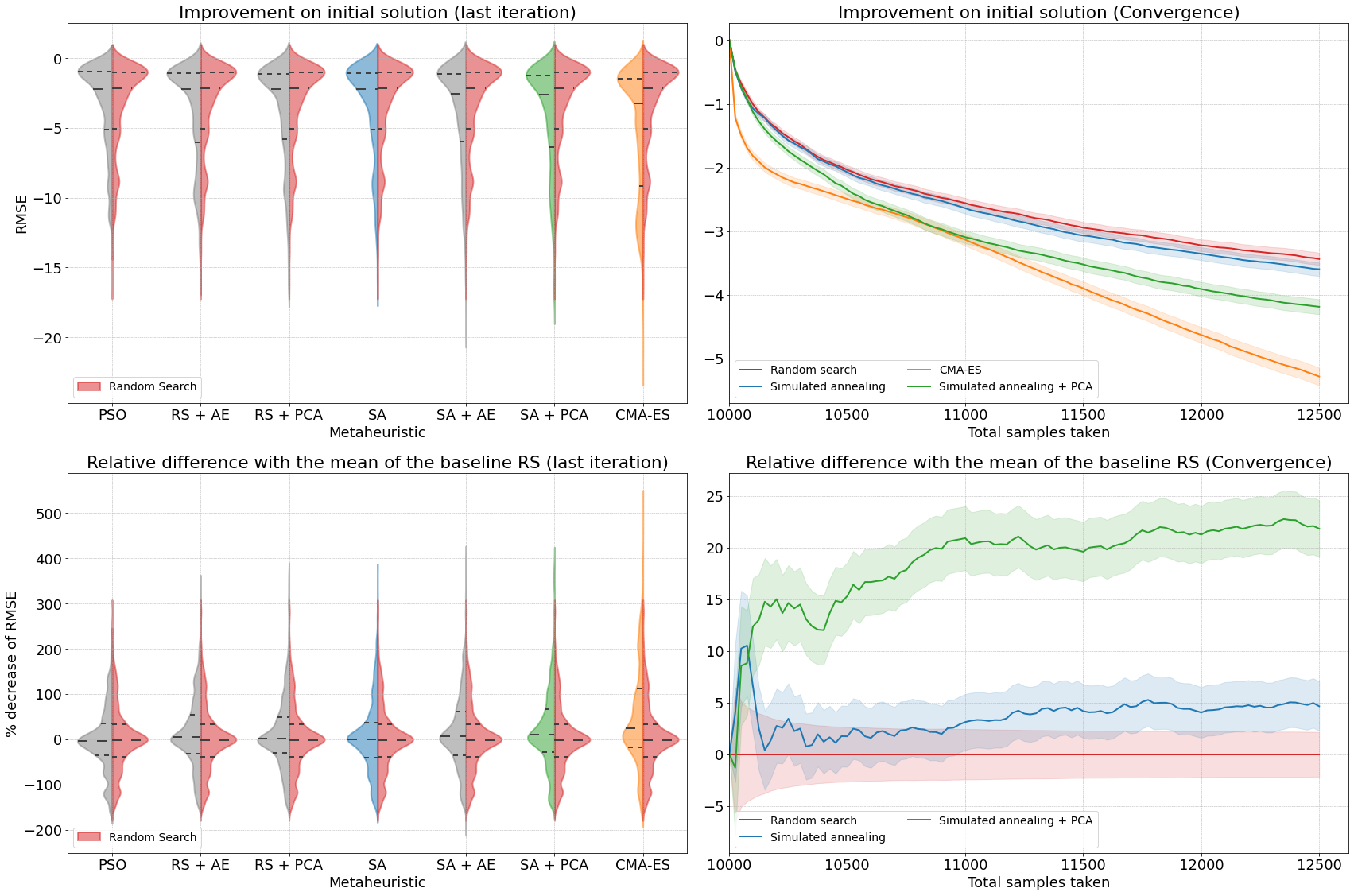}
  \caption{Performance of DR-FFIT-augmented search methods (Hodgkin-Huxley use-case). Same data as for Figure 5,  with full empirical distributions (left plots):  . Top row: (left) Empirical distributions of the performances obtained at the last iteration of the tested search methods ;  (right) the related average performances (shaded areas show standard-errors) as a function of the number of samples used per observation for 4 selected methods (right). Bottom row: (left) Empirical distributions of the fMRD obtained at the last iteration of the tested search methods; (right) average MRD (shaded areas show standard-errors) as a function of the number of samples used for 3 of the 4 selected methods. }
  \label{suppfigure2}
\end{suppfigure}

\end{document}